\documentclass[preprint,11pt,3p]{elsarticle}

\usepackage{amssymb}
\usepackage[usenames]{xcolor}

\usepackage{xspace}
\usepackage{subfig}
\usepackage[dvips]{epsfig}
\usepackage{epstopdf}
\usepackage{array}
\usepackage{framed}
\usepackage{comment}
\usepackage[normalem]{ulem}
\journal{Computer Physics Communications}

\newcommand{\parspace}{\vspace{0.3cm}}
\newcommand{\ispace}{\hspace*{0.6cm}}

\newenvironment{escom}{\parspace\noindent\ispace\ignorespaces}{\parspace

\noindent\ignorespacesafterend}

\newcommand{\es}{\mbox{\textsf{ESPResSo}}\xspace}

\newcommand{\var}[1]{\ensuremath{\mathrm{\mathit{#1}}}}

\usepackage{url}

\makeatletter
\def\url@leostyle{%
  \@ifundefined{selectfont}{\def\UrlFont{\sf}}{\def\UrlFont{\small\ttfamily}}}
\makeatother
\usepackage{upgreek}
\usepackage{amsmath}

\begin{document}

\begin{frontmatter}




\title{An \es implementation of elastic objects immersed in a fluid}


\author[uniza]{I. Cimr\'ak\footnote{The work of I. Cimr\'ak was supported by the Slovak Research and Development Agency under the contract No. APVV-0441-11, and by the Marie-Curie grant No. PCIG10-GA-2011-303580.}}
\author[fhstp]{M. Gusenbauer\footnote{M. Gusenbauer was supported by the N\" O Forschungs- und Bildungsges.m.b.H. 
(NFB) through the Life Science Calls.}}
\author[unizakis]{I. Jan\v cigov\'a\footnote{The work of I. Jan\v cigov\'a was supported by the Slovak Research and Development Agency under the contract No. APVV-0441-11.}}
\address[uniza]{Department of Software Technologies, Faculty of Management Science and Informatics, University of \v Zilina, Slovakia}
\address[fhstp]{St. Poelten University of Applied Sciences, Industrial Simulation, St. Poelten, Austria}
\address[unizakis]{Department of InfoCom Networks, Faculty of Management Science and Informatics, University of \v Zilina, Slovakia}

\begin{abstract}
We review the lattice-Boltzmann (LB) method coupled with the immersed boundary (IB) method for the description of combined flow of particulate suspensions with immersed elastic objects. We describe the implementation of the combined LB-IB method into the open-source package \es. We present easy-to-use structures used to model a closed object in a simulation package, the definition of its elastic properties, and the interaction between the fluid and the immersed object. We also present the test cases with short examples of the code explaining the functionality of the new package.
\end{abstract}


\end{frontmatter}

\def\hs{\hspace{0.0cm}}
{
\renewcommand{\arraystretch}{1.0}
\noindent
\begin{tabular}{p{3.2cm}p{11.5cm}}
\hline
\ 						& 																								\\
\multicolumn{2}{l}{\bf FRAMEWORK FEATURES }																				\\
\ 						& 																								\\
{\bf Feature} 			& {\bf OIF command example and options}															\\
\ 						&																								\\
{\bf Templates} 		& \verb oif_create_template \ \ \ \verb template-id \ 0											\\
\hs geometry			& \hs \verb nodes-file \ "n.dat", \verb triangles-file \ "t.dat"								\\
\hs elastic properties	& \hs \verb ks \ 0.1, \verb kb \ 0.03, \verb kal \ 0.1, \verb kag \ 0.4, \verb kv \ 1.0			\\
\end{tabular}

\vspace{0.3cm}
\noindent
\begin{tabular}{p{3.2cm}p{11.5cm}}
{\bf Objects} 			& \verb oif_add_object  \ \ \ \verb object-id \ 0	 											\\
\hs source template		& \hs \verb template-id \ 1																		\\
\hs position, rotation	& \hs \verb origin \ 10 8 13, \hs \verb rotate \ $\pi/2$ \ 0 \  $\pi/4$							\\
\end{tabular}

\vspace{0.3cm}
\noindent
\begin{tabular}{p{3.2cm}p{11.5cm}}
{\bf Properties}		& \verb oif_object_set  \ \ \ \verb object-id \ 0	 											\\
\hs position			& \hs \verb origin \ 10 8 13																	\\
\hs external forces		& \hs \verb force \ 3 0 0																		\\
\hs deformation			& \hs \verb mesh-nodes \ "deformed.dat"															\\
\end{tabular}

\vspace{0.3cm}
\noindent
\begin{tabular}{p{3.2cm}p{11.5cm}}
{\bf Output}			& \verb oif_object_output  \ \ \ \verb object-id \ 0 											\\
\hs vtk visualization	& \hs \verb vtk-pos \ "pos.vtk"	 									\\
\hs deformation			& \hs \verb mesh-nodes \ "deformed.dat"															\\
\end{tabular}

\vspace{0.3cm}
\noindent
\begin{tabular}{p{3.2cm}p{11.5cm}}
{\bf Mesh}				& \verb oif_check_mesh  							 											\\
\hs mesh files			& \hs \verb nodes-file \ "n.dat", \verb triangles-file \ "t.dat" \								\\
\hs mesh analysis		& \hs \verb orientation \ 																		\\
\hs mesh repair			& \hs \verb repair \ "r.dat"																	\\
\end{tabular}

\vspace{0.3cm}
\noindent
\begin{tabular}{p{3.2cm}p{11.5cm}}
{\bf Other}				&						   							 											\\
\hs initialization		& \verb oif_init \																				\\
\hs information			& \verb oif_info \																				\\
\ 						& 																								\\
\hline
\end{tabular}
}

\section{Introduction }
\label{sec:intro}

\begin{figure}[h]
   \centering
      \subfloat[]{\epsfig{file=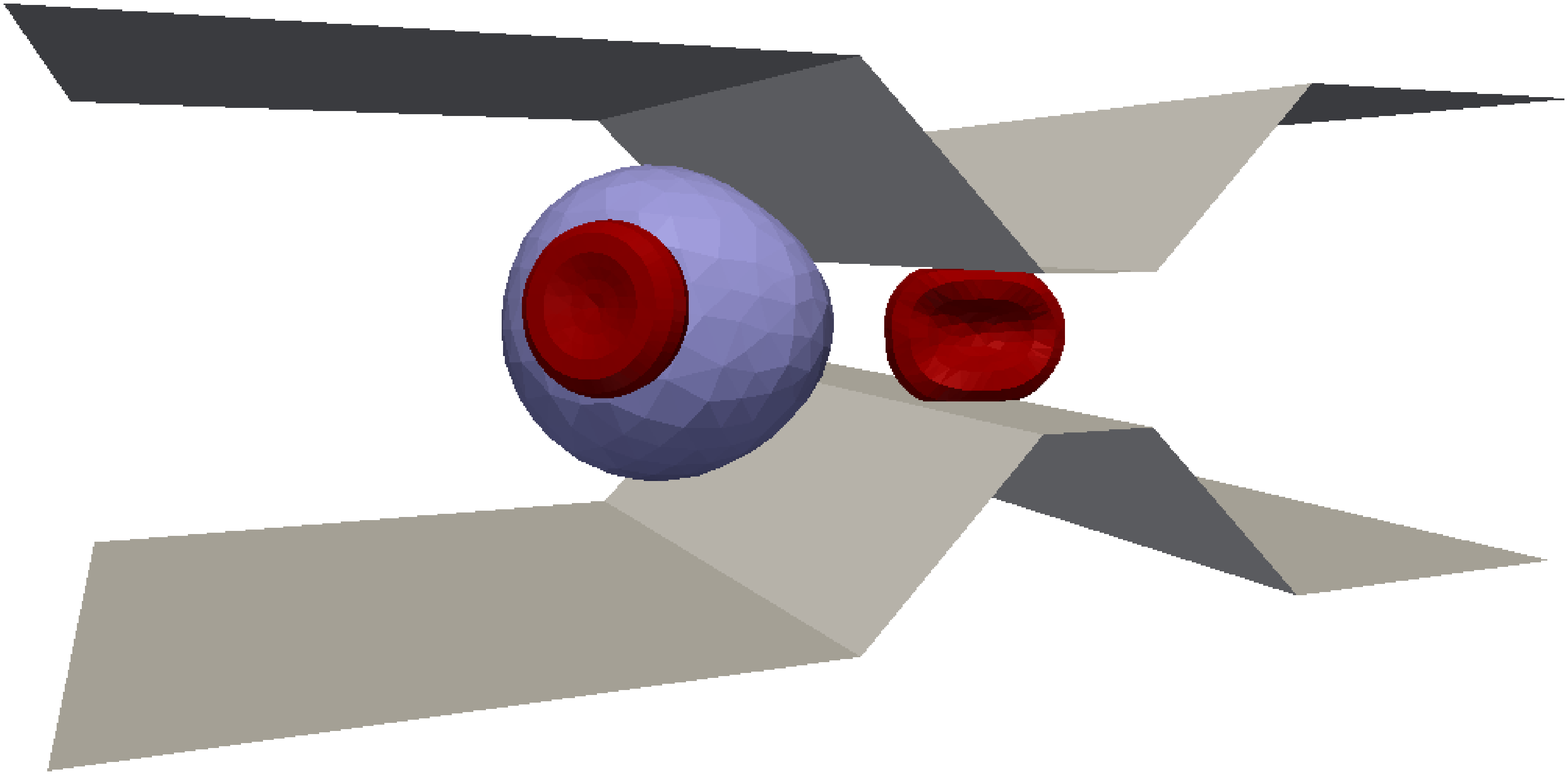,width=0.7\textwidth}}\qquad
      \subfloat[]{\epsfig{file=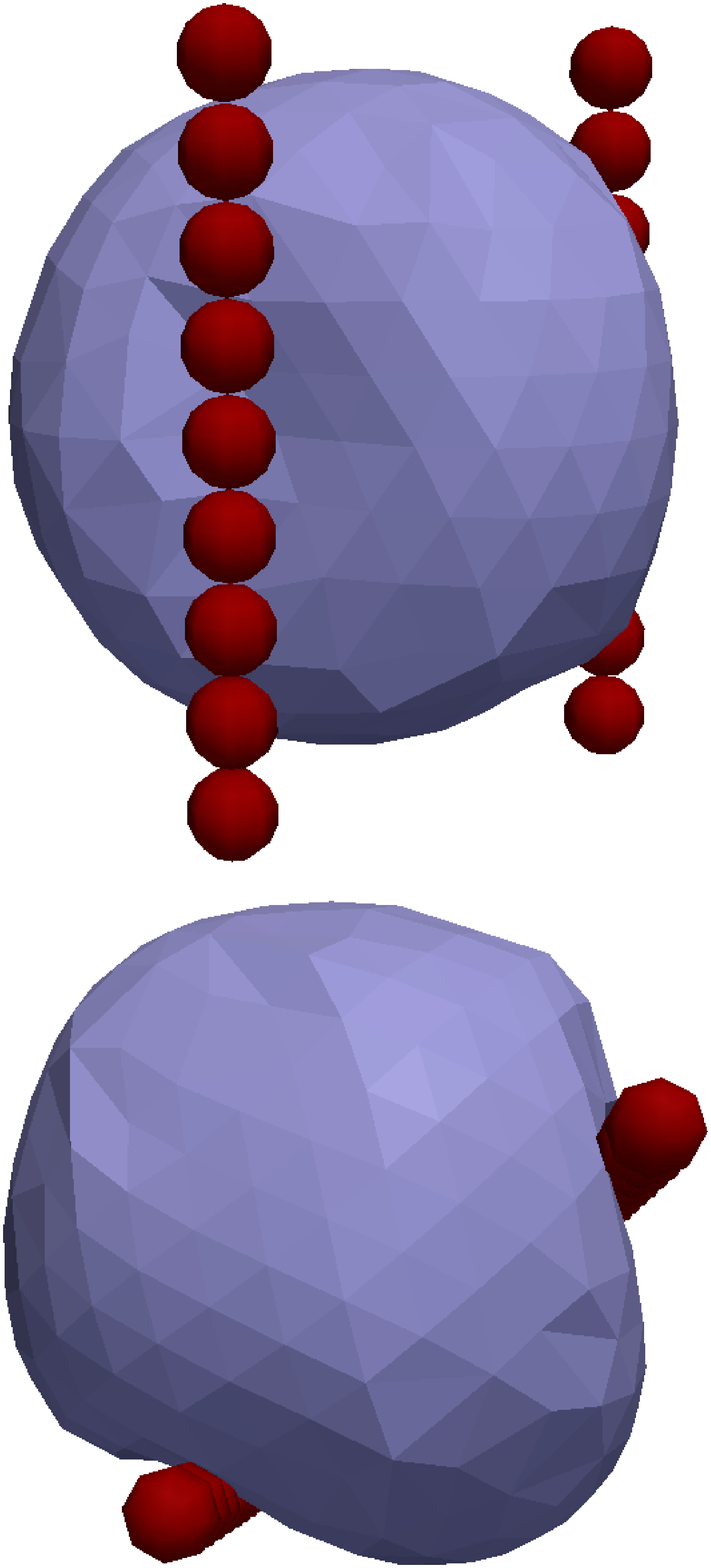,width=0.15\textwidth}}\qquad
  \caption{Elastic objects in biomedical applications a) Simulation of blood flow through tunable magneto-active polymer channel \cite{gusenbauer2013} 
    b) Circulating tumor cell captured in magnetic bead trap (side and top view) \cite{gusenbauer2012} }
  \label{blood:fig:simulations}
\end{figure}

\noindent The simulations of flow with immersed objects are a key technique for analysis in numerous areas. One can simulate the propulsion of bacteria \cite{maniyeri2012}, another application is the analysis of fish-like behavior \cite{gilmanov2005}. Our current research is motivated by biomedical applications for which simulation of blood on the level of blood cells immersed in a fluidic blood plasma is crucial (Fig. \ref{blood:fig:simulations}). These applications require understanding of processes depending on individual behavior of particular cells. Among others, we speak about microfluidic devices aimed for filtering of circulating tumor cells from blood \cite{chen2012}.

We utilize the lattice-Boltzmann (LB) method for the description of the fluid dynamics and the immersed boundary (IB) method for the description of the immersed objects. The coupling of both methods provides an accurate description of the complete fluid-structure interactions.

The lattice-Boltzmann method is a powerful technique enabling fast and accurate computation of the fluid flow in complex structures \cite{guo2002,kutay2006,geller2006}. In its nature, it is suitable for parallelization, performing very well in the benchmark examples \cite{kandhai1998,satofuka1999,williams2009}. The LB method describes the flow dynamics by giving information about the fluid velocity and pressure in each space-time discretization point of the computational domain.

The immersed boundary (IB) method is based on the discretization of the boundary of the immersed object. The method was first introduced by Charles S. Peskin in 1972 \cite{peskin1972}. The advantage of the method lies in its ability to capture the changes in the shape of the object without re-meshing and adaptive refinement of the underlying meshes. This is a crucial advantage for the simulation of moving boundaries.

\section{Method}\label{sec:method}

\noindent The \es package is aimed for soft-matter simulations. Originally, it was intended for simulations of different chemical systems where molecules and atoms are involved. These atoms are modeled as particles with their own mass. User defines bonds and potentials that generate forces between the particles. Consequently, particles move in the space subject to these forces and numerous other constraints.

In later releases of \es, the LB method was implemented to simulate soft-matter systems immersed in fluid. However, the package had no concept of a closed object. All the entities simulated in \es were open, tree-like structures without inner volume and surface. Our idea was to extend the functionality of \es by introducing closed objects with their own particle management in order to simulate elastic objects immersed in a fluid.

In this section, we first briefly describe the LB method as it was implemented in the original version of \es. Then we describe the IB method that we have implemented.
	
{\bf Lattice Boltzmann method:} This method is based on fictive particles. These particles perform consecutive propagations and collisions over a fixed discrete lattice. 
Consider a lattice placed over the three-dimensional domain and consisting of identical cubic cells. This lattice creates an Eulerian grid which is fixed during the entire simulation. The variable of interest in the LB method is $n_i(x, t)$ which is the particle density function for the lattice point $x$, discrete velocity vector $e_i$, and time $t$. We use the D3Q19 version of the LB method (three dimensions with 19 discrete directions $e_i$, so $i = 0, . . . , 18$). The governing equations in the presence of external forces, are 
\begin{equation}\label{eq:lbequation}
\underbrace{\vphantom{\frac{1}{\tau}}n_i(x+e_i\delta_t,t+\delta_t) = n_i(x,t)}_{propagation} - \underbrace{\vphantom{\frac{1}{\tau}}\Delta_i(\textbf{n}(x,t))}_{collision} + \underbrace{\vphantom{\frac{1}{\tau}}f_i(x,t)}_{external forces}
\end{equation}
where $\delta_t$ is the time step and $\Delta_i$ denotes the collision operator that accounts for the difference between pre- and post-collision states and satisfies the constraints of mass and momentum conservation. $f_i$ is the external force exerted on the fluid. We refer to \cite{Ahlrichs1998} and \cite{dunweg2008} for details on the LBM. The macroscopic quantities such as velocity $u$ and density $\rho$ are evaluated from
\begin{displaymath}
\rho(x,t) = \sum_{i}n_i(x,t)\qquad\mathrm{and}\qquad \rho(x,t)u = \sum_i n_i(x,t)e_i.
\end{displaymath}


{\bf Immersed Boundary method:} This method is based on the discretization of object's boundary. The boundary is covered by a set of IB points linked by triangular mesh, which is called Lagrangian mesh. The positions of IB points are not restricted to any lattice. To take the mechano-elastic properties of the immersed objects into account, geometrical entities in this mesh (edges, faces, angles between two faces, \ldots ) are used to model stretching, bending, stiffness, and other properties of the boundary. They define forces according to the current shape of the immersed object that are exerted on IB points. These forces cause motion of the IB points according to the Newton's equation of motion
\begin{equation}\label{eq:newton}
m\frac{d^2 X_j}{dt^2} = F_j
\end{equation}
where $m$ is the mass of the IB point, $X_j$ is the position and $F_j$ is the force exerted on the particular IB point $j$. The source of $F_j$ is either from the above mentioned elasto-mechanical properties of the immersed object, or from the fluid-structure interaction. 

Equations (\ref{eq:lbequation}) and (\ref{eq:newton}) describe two different model components on two different meshes: the motion of the fluid and the motion of the immersed objects. For coupling, \es uses an approach from \cite{Ahlrichs1998} with a drag force that is exerted on a sphere moving in the fluid. Analogous to the Stokes formula for the sphere in a viscous fluid, we assume the force exerted by the fluid on one IB point to be proportional to the difference of the velocity $v$ of the IB point and the fluid velocity $u$ at the same position,
\begin{equation}\label{eq:dragforce}
F_j = \xi (v - u).
\end{equation}
Here $\xi$ is a proportionality coefficient which we will refer to as the friction coefficient. In the previous expression, the velocities $v$ and $u$ are computed at the same spatial location, whereas we posses $u$ in fixed Eulerian grid points and $v$ in moving Lagrangian IB points. Therefore, for computation of $u$ at the IB point, we use linear interpolation of the $u$ values from nearby fixed grid points.

There is also an opposite effect: not only fluid acts on the IB point, but also the IB point acts on the fluid. Therefore, the opposite force $-F_j$ needs to be transferred back to the fluid. $F_j$ from the location of the IB point is distributed to the nearest grid points. Distribution is inversely proportional to the cuboidal volumes with opposite corners being the IB point and the grid point \cite{Ahlrichs1998}.


\section{Object-in-fluid framework}
\label{sec:implementation}


\noindent We implement an object-in-fluid (OIF) framework into \es that allows the use of objects with inner volume, for example blood cells, magnetic beads, capsules and so on. We use LB-IB method where IB points from the triangulation are identified with \es \emph{particles}. The edges of the triangular mesh are identified as bonds in \es. Bonds generate elastic forces that keep the shape of the object. The movement of the object is achieved by applying calculated forces to the IB points.


The immersed objects are composed of a membrane encapsulating the fluid inside the object. For now, the inside fluid must have the same density and viscosity as the outside fluid. 




The original \es works with particles that can be defined by a single \verb part \ command executed for each particle. Particles are virtually connected with bonds executing \verb inter \ and \verb part \ \verb bond \ commands. To create a mesh, e.g. a triangular mesh with 300 mesh nodes covering a sphere, one needs to execute a series of commands analyzing the mesh, computing hundreds of edge lengths, all angles between triangles, etc. Afterwards, more than thousand commands must be executed to add the particles and define the bonds. In the OIF framework we enable the automatization of this process with numerous features. 








The OIF framework works  in two steps. First, one needs to create a template, then one can add actual objects. The template has its own elastic properties and shape. Templates are not real objects, they only keep information about the objects to be created. After  a template (or several templates) is created, one can add actual objects to  the simulation  by specifying the template from which the object takes its geometry and elastic properties, the location and, optionally, the rotation of the object. 

During the simulation, one can analyze the objects by printing their physical properties. Also, different quantities characterizing the objects can be set.

Further functionality includes analysis of the mesh files, output for the VTK visualization using a standalone visualization software, e.g. ParaView \cite{paraview}. Detailed description of all currently implemented features of the OIF framework is in Section \ref{oif_features}. However, the OIF framework is a lively project that is being extended on regular base. Therefore for up-to-date documentation on the latest developments we recommend the \es documentation \cite{espressoDocumentation}.


\section{Elastic interactions}
\label{sec:inter-bonded-fsi}

\noindent The generic implementation of bonded interactions in \es is described in detail in \cite{Limbach2006,Arnold2013}. 
Bonded interactions are always defined between two, three or sometimes four points. They generate forces that are exerted
on each IB point. These forces must be evaluated at a certain point of the integration algorithm. At that moment, with given positions of the IB points, the forces for each specific interaction can be calculated and added to the IB points. 


\vspace{5pt}
\noindent
{\bf Parameters:} 
There are several parameters involved in this model. All of them should be calibrated with respect to the application. 
\begin{description}
\item[] \textit{Mass of the particles:} Every particle has its mass that influences the dynamics. Its default value is one, however it can be changed when adding the object into simulation with \verb oif_add_object \ \verb mass \ command.
\item[] \textit{Friction coefficient $\xi$:} The parameter describing the strength of fluid-particle interaction is the \verb friction \ parameter from the original \es command \verb lbfluid , see (\ref{eq:dragforce}).
\item[] \textit{Parameters of elastic moduli:} Elastic behavior can be described by five different elastic moduli: stretching, bending, local
and global area preservation and volume preservation. Each of them has its own scaling parameter: \verb ks, \verb kb, \verb kal, \verb kag, \verb kv. \ 
\end{description}
The mass of the particles and the friction coefficient can be calibrated using the known analytical values of the drag coefficients for ellipsoidal objects.
More details about the calibration is given in \cite{cimrak11a}.

The elastic parameters depend on the way how the physical properties are modeled. For each physical property there are always numerous approaches how to model that particular property. As a very simple example let us discuss the local area conservation. Assume an elastic object has a relaxed shape, in which this objects remains unless external forces are applied. In the relaxed shape, each local part of the surface has its area. The general idea is that once the local part of the object's surface has a larger area than in the relaxed state then this part must be shrunk. If the current area is smaller than that in the relaxed shape, the surface must be expanded. 

To implement this general idea we show two different approaches. Both approaches first compute the magnitude of the force that will be applied to each vertex of the triangles in the mesh. This magnitude is proportional to the difference between the current area of the triangle and the area in the relaxed shape. The directions of the forces however differ in the two approaches. First approach puts the forces in the direction from the vertex to the centroid of the triangle. The second approach applies the forces in the vertices in the direction of the respective altitudes of the triangle. Both approaches follow the main idea, however the actual effect on the triangle is different.

In the current version of the OIF framework, we work with the elastic models taken from \cite{Dao2003}. More details about the mechanical and biological aspects specifically for red blood cells are presented in the cited work. The proper calibration to fit the experimental data has been performed in \cite{cimrak11a,Cimrak2013}.

However, we are aware of the different approaches. Authors in \cite{Fedosov2010a, Fedosov2010b, Odenthal2013} use the energy approach where the forces are derived from the energy contributions rather than defined explicitly as in \cite{Dao2003}. The OIF framework is flexible enough to implement any method based on the application of forces to the nodes in the triangular mesh.

The following interactions are implemented in order to mimic the mechanics of elastic or rigid objects immersed in the LB fluid flow. They are all based on the fact that an elastic object without any external forces has a relaxed shape. The object always tends to recover this relaxed shape.  Once this shape is deformed, elastic forces act on the boundary of the object, trying to return it to its relaxed state. So for example, all edges of the triangulation in the relaxed shape have a reference length which we call relaxed length. Similarly, there is a relaxed angle between two incident triangles in the mesh, a relaxed area of each triangle in the mesh and so on.

The elastic forces have specific expressions for each elastic modulus. Their mathematical formulations have been taken from \cite{Dupin2007}. 

\begin{figure}[h]
\begin{center}
  \includegraphics[width=.9 \textwidth]{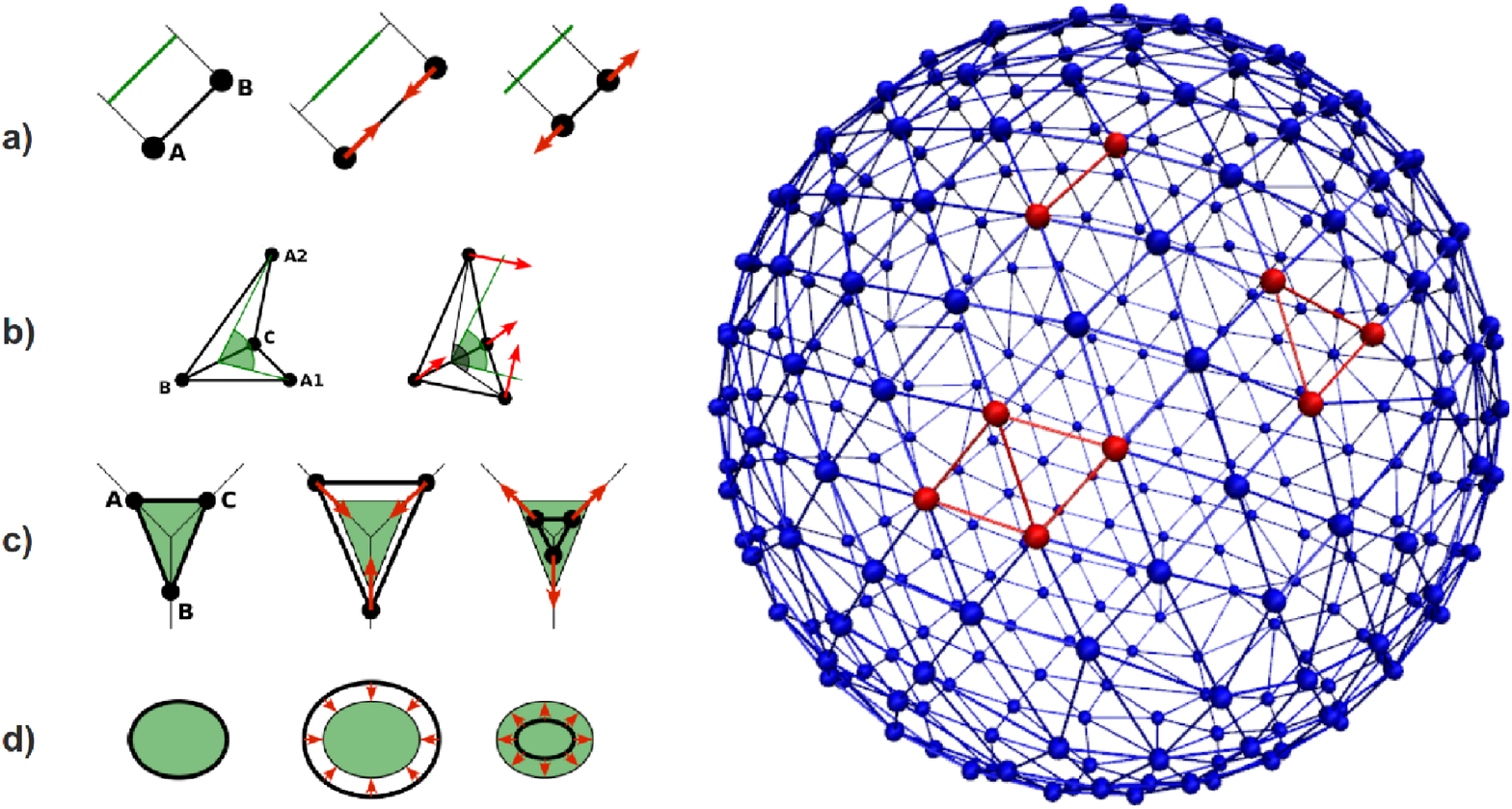}
\end{center}
\caption{Elastic sphere model in \es: \newline a) stretching force, b) bending force, c) local area force, d) volume force/global area force}\label{fig:model}
\end{figure}

\subsection*{Stretching force}
\parspace



\noindent This type of interaction is available for closed 3D immersed objects as well as for 2D sheet flowing in the 3D flow. 

It forces each triangulation edge to adjust towards its relaxed length. For each edge of the mesh, $L$ is the current length of the edge. By $L^0$ we denote the relaxed length. In the case of $L = L^0$ no forces are added. $\Delta L = L - L^0$ denotes the deviation from the relaxed length. The stretching force acting on edge endpoints is computed using 

\begin{equation}\label{stretching}
F_\mathrm{s} = k_\mathrm{s}\kappa(\lambda)\frac{\Delta L}
{L^0}n.
\end{equation}

\noindent Here, \var{n} is the unit vector pointing from the one edge endpoint to the other, $k_\mathrm{s}$ is the stretching constant, $\lambda = L/L^0$, and $\kappa$ is a nonlinear function that resembles neo-Hookean behavior

\begin{equation}
\kappa(\lambda) = \frac{\lambda^{0.5} + \lambda^{-2.5}}
{\lambda + \lambda^{-3}}.
\end{equation}

\noindent For linear behavior of the stretching force we simply have $\kappa = 1$.
The stretching force acts between two particles and is symmetric (Figure \ref{fig:model}a).


\subsection*{Bending force}
\parspace



\noindent Beside the stretching force, the tendency of an elastic object to maintain the relaxed shape is governed also by prescribing the preferred angles between two incident triangles of the mesh. This type of interaction is available for closed 3D immersed objects, as well as for 2D sheet flowing in the 3D flow.

The angle between two triangles in the resting shape is denoted by $\theta^0$. For closed immersed objects, one always has to set the inner angle. The deviation of this angle $\Delta \theta = \theta - \theta^0$ is computed and defines the bending force for each triangle. 
\begin{equation}
F_\mathrm{b} = k_\mathrm{b}\frac{\Delta \theta}{\theta^0} n
\end{equation}

\noindent Here, $n$ is the unit normal vector to the triangle and $k_\mathrm{b}$ is the bending constant. The force $F_\mathrm{b}$ is assigned to the two vertices not belonging to the common edge. The opposite force divided by two is assigned to the two vertices lying on the common edge (Figure \ref{fig:model}b). 

Unlike the stretching force, the bending force is strictly asymmetric. For more details, see the \es documentation \cite{espressoDocumentation}.
Notice, that concave objects can be defined as well. If $\theta_0$ is larger than $\pi$, then the inner angle is concave.







\subsection*{Local area conservation}
\parspace

  

\noindent This interaction conserves the area of the triangles in the triangulation. This type of interaction is available for closed 3D immersed objects, as well as for 2D sheet flowing in the 3D flow.

The deviation of the triangle surface $S$ is computed from the triangle surface in the resting shape $\Delta S = S - S^0$. The area constraint assigns the following shrinking/expanding force to every vertex 

\begin{equation}
F_\mathrm{al} = -k_\mathrm{al}\frac{\Delta S}{S}w
\end{equation}

\noindent where $k_\mathrm{al}$  is the area constraint coefficient, and $w$ is the unit vector pointing from the centroid of triangle to the vertex. Similarly, analogical forces are assigned to other vertices of the triangle (Figure \ref{fig:model}c). This interaction is symmetric.




\subsection*{Global area conservation}
\parspace



\noindent This type of interaction is available for closed 3D immersed objects, as well as for 2D sheet flowing in the 3D flow.

The conservation of local area is sometimes too restrictive. The current surface of the whole immersed object is denoted by \var{S_g}, the surface in the relaxed shape by \var{S_g^0} and $\Delta \var{S_g} = \var{S_g} - \var{S_g^0}$. The global area conservation force is then defined as

\begin{equation}
F_\mathrm{ag} = - k_\mathrm{ag}\frac{\Delta S_g}{S_g}w
\end{equation}

\noindent The force $F_\mathrm{ag}$ is assigned to all vertices of the object (Figure \ref{fig:model}d).

Unlike the previous three forces (and related bonds), this one is different from classical bonds in \es, because it requires an advance calculation on the scale of the whole immersed object - to get its total area. This calculation has been implemented to allow for proper area computation also in parallelized simulations when parts of the given object lie in different blocks of the decomposed domain. In particular, consider a case when the immersed object lies in several regions that belong to different computational nodes. To compute the surface of such immersed object, each computational node needs to compute partial surface of the corresponding part and send the information to master to gather the whole surface of the object. To implement this, at each surface computation, a message is broadcasted to all computational nodes that adds the partial surface of the object. After gathering the information from all nodes, we can compute the actual forces for the global area conservation.



\subsection*{Volume conservation}
\parspace



\noindent This type of interaction is available solely for closed 3D immersed objects.

The deviation of the object's volume $V$ is computed from the volume in the resting shape $\Delta V = V - V^0$. For each triangle the following force is computed

\begin{equation}
F_\mathrm{v} = -k_\mathrm{v}\frac{\Delta V}{V^0} S\ n
\end{equation}

\noindent where $S$ is the area of triangle, $n$ is the normal unit vector of triangle's plane, and $k_\mathrm{v}$ is the volume constraint coefficient. The volume of the immersed object is computed from

\begin{equation}
V = \sum_{triangles}S\ n\cdot h
\end{equation}

\noindent where the sum is computed over all triangles of the mesh and $h$ is the unit vector from the centroid of triangle to a plane which does not cross the object. The force $F_\mathrm{v}$ is equally distributed to all three vertices (Figure \ref{fig:model}d).

Similar to the computation of global area force, the volume force also requires a prior calculation on the scale of the whole object - to get the total volume. And again, it is implemented in such a way that partial volumes are added together in parallelized simulations when the object belongs to more than one block of decomposed domain.

\subsection*{Remark}
The actual values of $k_\mathrm{s},k_\mathrm{b},k_\mathrm{al},k_\mathrm{ag},k_\mathrm{v}$ need to be calibrated for each type of the elastic object separately. They strongly depend on the physical values of elastic moduli. Their values can be determined e.g. from stretching experiments performed with red blood cells \cite{Dao2003}. The stretching force expression (\ref{stretching}) yields the sensitivity of  $k_\mathrm{s}$ on the density of the triangular mesh. This dependence was elaborated in \cite{Cimrak2013}.







\def\descriptionsection #1{
\vspace{5pt}

\noindent
{\bf #1}

}

\section{Implementation details and OIF command description}
\label{oif_features}
\parspace
\noindent All commands and variables in the OIF framework start with prefix \verb oif_ . Syntax in the OIF framework follows the structure from \es. Each OIF command consists of command keyword followed by several option - arguments couples. The order of these couples is arbitrary. Some options of the commands are mandatory, some are optional. The options mostly have one or more arguments but there are also options without arguments.

The mandatory options are listed directly after the command keyword, the optional ones are listed inside brackets.
\descriptionsection{Initialization, general info}
\nopagebreak
\begin{escom}
\begin{tabular}{ll}
\verb oif_init \ 	& \\
\verb oif_info \ 	& \\
\end{tabular}
\end{escom}
\noindent
The first command initializes the OIF framework and must be used before any other OIF commands. It creates global variables used in the OIF framework. These variables are accessible on the Tcl level in \es, but for the ordinary user of the OIF framework it is not necessary to access them. The detailed description of the OIF variables is in the \es documentation.

The second command prints basic information about the current number of created templates and objects, etc. 

Both commands are without options.

\def\ssp{\ \ \ }
\descriptionsection{Template creation}
\nopagebreak
\begin{escom}
\begin{tabular}{ll}
\verb oif_create_template \ 	& \verb template-id \ \textit{tid} \ \verb nodes-file \ \textit{nfile} 			\\
								& \verb triangles-file \ \textit{tfile} \ [\verb stretch \ \textit{sx sy sz}] 		\\
								& [\verb ks \ \textit{val}] \ [\verb kb \ \textit{val}] \ [\verb kal \ \textit{val}] \ [\verb kag \ \textit{val}] \ [\verb kv \ \textit{val}] 								\\
\end{tabular}
\end{escom}
\begin{description}
\item[] \verb template-id \ 
Each template must have its unique identifier \textit{tid}. This identifier starts from 0 and must be numbered consecutively.
\item[] \verb nodes-file , \verb triangles-file \ 
The information about the geometry is extracted from a surface triangulation called mesh. Spatial positions of the mesh points are obtained from the \textit{nfile}. This simple text file contains lines and each line has three spatial coordinates of a mesh point. These coordinates should be centered in such a way that the approximate center of the object is located at the origin of the coordinate system. The order of mesh points in \textit{nfile} defines the IDs of the points starting from 0. Further, a file \textit{tfile} contains lines and each line has three integer numbers representing the IDs of three mesh points in one triangular face of the mesh. It is crucial that three IDs in one line of \textit{tfile} are in the right order. They must fulfill the rule that the normal vector of the mesh triangle must point inside the object. Specifically, assume A, B and C are three points of a triangle in the mesh and their IDs are 3, 5 and 8. Compute the cross product $n = AB \times AC$. If the resulting vector $n$ 
(which is perpendicular to the triangle $ABC$) points inside the object then we have line 3 5 8 in \textit{tfile}. If the vector $n$ points outside the object, we change the order of B and C and we have line 3 8 5 in \textit{tfile}.  
\item[] \verb stretch \ 
The geometry described in \textit{nfile} and \textit{tfile} can be customized by stretching of the coordinates. This option will stretch x-coordinates by factor \textit{sx}, y-coordinates by factor \textit{sy}, and z-coordinates by factor \textit{sz}.
\item[] \verb ks , \verb kb , \verb kal , \verb kag , \verb kv \ 
These options define the prefactors for elastic forces. 
\begin{itemize}
\item \verb ks \ssp sets the stiffness of the spring between two particles on each edge of triangulation mesh. This parameter controls the stretching of the object.
\item \verb kb \ssp  sets the stiffness of the spring between two triangles with common edge. This parameter controls the bending of the object.
\item \verb kal \ssp  controls the conservation of the area of triangles.
\item \verb kag \ssp  controls conservation of the total surface area of the immersed object.
\item \verb kv \ssp  controls the conservation of the total volume of the immersed object.
\end{itemize}
More detailed explanation on different types of elastic forces is in Section \ref{sec:inter-bonded-fsi}.
\end{description}

\descriptionsection{Adding of the object into simulation}
\nopagebreak
\begin{escom}
\begin{tabular}{ll}
\verb oif_add_object \ 	& \verb object-id \ \textit{oid}  \ \verb template-id \ \textit{tid} \ \verb origin \ \textit{ox oy oz}	\\
						& \verb part-type \ \textit{ptype} \ [\verb rotate \ \textit{rx ry rz}] \ [\verb mass \ \textit{val}]		\\
\end{tabular}
\end{escom}
\begin{description}
\item[] \verb object-id \ 
Each object must have its unique identifier \textit{oid}. This identifier starts from 0 and must be numbered consecutively.
\item[] \verb template-id \ 
The geometry and elastic properties of each object are copied from a template identified by \textit{tid}.
\item[] \verb origin \ 
The object is placed in the space such that the approximate center of the object has coordinates (\textit{ox, oy, oz}).
\item[] \verb part-type \ 
In \es, every particle has its type and this type is used for interactions, e.g. with  the walls, etc. In the OIF framework, all particles that form one object have the same part-type \textit{ptype}. Note, that two different objects may have the same part-type.
\item[] \verb rotate \ 
Each object can be rotated. This does not influence the geometry and the elastic properties. The object will be rotated by $rx$ radians around the x axis. Similarly for \textit{ry} and \textit{rz}.
\item[] \verb mass \  
Sets the mass of particles for this object.
\end{description}

\descriptionsection{Visualization and data output}
\nopagebreak
\begin{escom}
\begin{tabular}{ll}
\verb oif_object_output \ 	& \verb object-id \ \textit{oid}  \ [\verb vtk-pos \ \textit{pfile} ] \\
							& [\verb mesh-nodes \ \textit{nfile} ] \\
\end{tabular}
\end{escom}
\begin{description}
\item[] \verb object-id \ 
Identifies the object under the consideration. 
\item[] \verb vtk-pos \ 
Outputs the triangular mesh of the object into the VTK-file \textit{pfile}. This file is directly readable e.g. with ParaView \cite{paraview} visualization software.
\item[] \verb mesh-nodes \ 
Outputs the current shape of the object into the file \textit{nfile}. This file can be used as an input for OIF command \verb oif_object_set , option \verb mesh-nodes . This option is used to start simulation from some deformed shape.
\end{description}

\descriptionsection{Object analysis}
\nopagebreak
\begin{escom}
\begin{tabular}{ll}
\verb oif_object_analyze \ 	& \verb object-id \ \textit{oid}  \ [\verb origin ] \ [\verb pos-bounds \ \textit{name}] 		\\
							& [\verb volume ]	\ [\verb surface-area ] \  [\verb velocity ]\\
\end{tabular}
\end{escom}
\begin{description}
\item[] \verb object-id \ 
Identifies the object under the consideration. 
\item[] \verb origin \ 
Prints the coordinates of the object's center. Since the object can be deformed, the coordinates of the origin are computed as an average of coordinates of all mesh points. 
\item[] \verb pos-bounds \ 
Computes six extremal coordinates of the object. More precisely, runs through all the mesh points and remembers the minimal and maximal x-coordinate, y-coordinate and z-coordinate. If \textit{b-name} is one of these: "z-min, z-max, x-min, x-max, y-min, y-max", then the procedure returns the corresponding value. If \textit{b-name} is "all", then the procedure returns a list of six numbers.
\item[] \verb volume \ 
Returns the volume of the object. 
\item[] \verb surface-area \ 
Returns the surface area of the object. 
\item[] \verb velocity \ 
Returns the velocity of the object computed as an average over the velocities of all mesh points.
\end{description}

\descriptionsection{Object customization}
\nopagebreak
\begin{escom}
\begin{tabular}{ll}
\verb oif_object_set	 \ 	& \verb object-id \ \textit{oid}  \ [\verb force \ \textit{fx fy fz}] \ [\verb origin \ \textit{ox oy oz}]\\
							& [\verb mesh-nodes \ \textit{nfile}] \\
\end{tabular}
\end{escom}
\begin{description}
\item[] \verb object-id \ 
Identifies the object under the consideration. 
\item[] \verb force \ 
Sets the force vector (\textit{fx,fy,fz}) to all mesh nodes of the object. Setting is done executing \es command \verb part \ \textit{i} \verb set \ \verb ext_force \ \textit{fx fy fz} for each particle of the object. 
\item[] \verb origin \ 
The object is translated so that its center coincides with the spatial point (\textit{ox,oy,oz}).
\item[] \verb mesh-nodes \ 
Providing the file \textit{nfile} one can change the current shape of the object.
\end{description}

\descriptionsection{Mesh analysis}
Sometimes it happens that the orientations of the triangles in \textit{tfile} are not correct. To check whether one has the correct orientations or to fix some wrong ones, the following OIF command can be used.

\begin{escom}
\begin{tabular}{ll}
\verb oif_mesh_analyze \ 	& \verb nodes-file \ \textit{nfile}  \ \verb triangles-file \ \textit{tfile} \\
							& [\verb orientation ]	\ [\verb repair \ \textit{rfile}] \ [\verb flip \ \textit{rfile}]\\
\end{tabular}
\end{escom}
\begin{description}
\item[] \verb nodes-file, \verb triangles-file \ 
The files that need to be checked or repaired.
\item \verb orientation \ 
Checks whether all triangles have correct orientation.
\item \verb repair \ 
Checks each triangle and corrects its orientation when necessary. Outputs the corrected incidence file \textit{rfile}. Note that this works for the convex objects and for some concave objects.
\item \verb flip \ 
Flips the orientation of all triangles and outputs new incidence file \textit{rfile}.
\end{description}

\vspace{1cm}
This description of the OIF framework is not intended to be complete. There are different demands on extension of the code and the current version included in the latest release of \es is always described in the user guide available at \cite{espressoDocumentation}.

\section{Example: simulation blood cells flow}
\label{sec:bloodcells}

The following example is part of our research focus in biomedical applications. Isolation of circulating tumor cells using antibody covered surfaces requires a high contact probability of blood cells and the activated surfaces (see \cite{gusenbauer2013,gusenbauer2012} for details). In microfluidics, the blood flow is usually laminar, therefore cells follow the streamlines, without even touching the surfaces. Obstacles in the channel help increase the contact with blood cells. 

In Table \ref{tab:script}, an example of a simple script is presented. The intention is to show that with a very simple script one can simulate quite complex physical system of the blood cells flowing inside a microfluidic channel. 

For the geometry of a red blood cell we use a triangulation with 400 nodes and 796 triangles. Triangulation is given in files \verb n.dat  and \verb t.dat.  First couple of lines from each file are presented in Table \ref{tab:files}.
\begin{table}
\begin{framed}
{\scriptsize
\begin{verbatim}
n.dat:                                  t.dat:
1.00000  0.00000  0.00017               9 17 11 
0.99262  0.00000  0.12129               18 10 13
0.99266  0.00000  -0.12096              1 9 5
0.95106  0.16995  0.00000               10 2 7
0.95106  -0.16995 0.00000               0 1 3
0.94356  0.16995  0.11920               2 0 3
\end{verbatim}
}
\end{framed}
\caption{Example of geometry files.}\label{tab:files}
\end{table}

\begin{table}
\begin{framed}
{\scriptsize
\begin{verbatim}
 1| setmd time_step 0.1        #
 2| setmd skin 0.2             # mandatory ESPResSo commands
 3| thermostat off             #
 4| setmd box_l 50 10 70       # simulation box is defined
 5| 
 6| oif_init                                              # 
 7| oif_create_template template-id 0 nodes-file "n.dat"\ #
 8|                     triangles-file "t.dat" \          # 
 9|                     ks 0.05 kb 0.01 kv 10.0           # template is created
10| oif_object_add object-id 0 template-id 0 \            #
11|                origin 5 3 15 part-type 0              #
12| oif_object_add object-id 1 template-id 0 \            #
13|                part-type 1 origin 5 3 23              # two objects are added
14| 
15| source boundaries.tcl                      # source file with definition 
16|                                            # of channel boundaries
17| inter 0 10 soft-sphere 0.0001 1.2 0.1 0.0  #  
18| inter 1 10 soft-sphere 0.0001 1.2 0.1 0.0  # 
19| inter 0 1 soft-sphere 0.005 2.0 0.3 0.0    # the boundaries are initialized
20| 
21| lbfluid grid 1 dens 1 visc 1.5 \      #
22|         tau 0.1 friction 0.5          # the fluid is initialized
23|       
24| set loop 0                                 # main iteration loop
25| while {$loop < 100000} {                   # 
26|   for {set i 1} {$i < 49} {incr i} {       #
27|     for {set j 1} {$j < 9} {incr j} {      # velocity of the fluid
28|       lbnode $i $j 1 set u 0.0 0.0 0.01    # is set on the inflow
29|     }                                      #
30|   }                                        #
31|   integrate 1                              # integration step
32|   incr loop                                #
33| }
\end{verbatim}
}
\end{framed}
\caption{Tcl script with simulation of flow of two red blood cells in a microfluidic channel.}\label{tab:script}
\end{table}

\begin{figure}[h]
\begin{center}
  \includegraphics[width=.9 \textwidth]{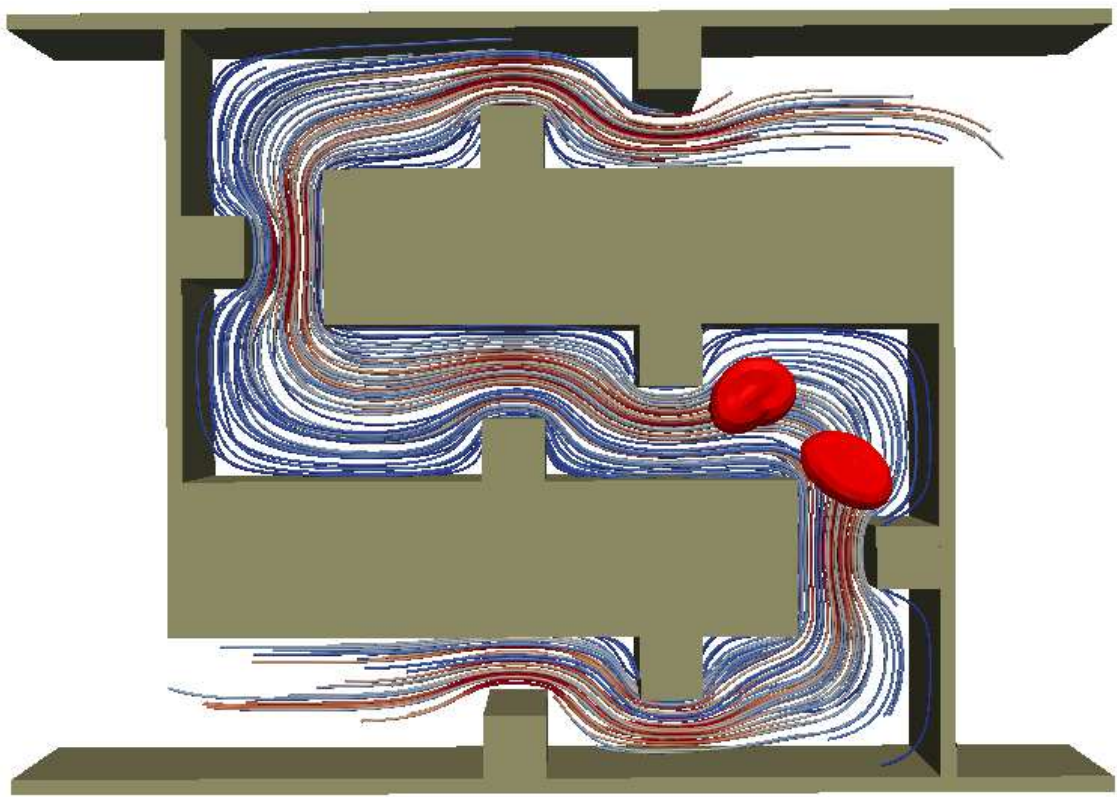}
\end{center}
\caption{Two red blood cells passing an S-shaped microfluidic channel with obstacles. Fluidic streamlines show the increased velocity at narrow gaps (red). Simulation data produced by \es is visualized with ParaView \cite{paraview}.}\label{fig:channelS}
\end{figure}

\begin{description}
\item[] {\it Initialization, lines 1--4.} 
In the first three lines of the script, several mandatory arguments for \es are set. In general, our calculations have been done with keeping the \verb time_step \ (line 1) for IB calculations equal to the \verb tau \ (line 22) for fluid propagation. Even though it is possible to thermalize particles in \es (using the \verb thermostat \ command, line 3), this is not relevant for simulations of objects on the scale of micrometers (e.g. blood cells), where the IB particles are not subject to thermal fluctuations.  In line 4 the dimensions of the simulation box are given. 
\item[] {\it OIF framework. lines 6--13.} 
Then we initialize the OIF framework with \verb oif_init  command. Further we create the template with template-id 0 by providing two mesh files and defining the elastic parameters. In this case we use the stretching, bending and volume conservation. The values for the elasticity parameters need to be determined from the modeled application. 

Further, we add two objects, created from template 0. We put them on different spatial locations specifying the coordinates of the origin. We set different part-type for both objects so that we can set the collision detection between them. Note, that the parameters can be given in any order.
\item[] {\it Boundaries, lines 15--19.} 
In line 15, we include an external Tcl script with definition of boundaries. We create a three-dimensional microfluidic channel with obstacles using the original \es commands \verb constraint \ \verb rhomboid, \ and \verb lbboundary \ \verb rhomboid, \ see Figure \ref{fig:channelS}. The external script contains standard Tcl and \es commands.

Further, in lines 17--19 we set the repulsive interactions to simulate object-object and object-wall collisions. 
\item[] {\it Fluid init and main loop, lines 21--33.} 
In lines 21--22 we initialize the fluid in the simulation. It is a standard \es command. Next we run the main loop. In each iteration, we first set the velocity of the fluid on the inflow to be constant (lines 26--30) using standard \es command \verb lbnode  and then we execute the integration procedure with \es command \verb integrate. Currently, we are exploring the option to set the fluid velocity using "moving" boundaries, e.g. a \verb lbboundary \ \verb rhomboid \ that performs a modified bounce back boundary condition for the fluid. This way of prescribing the fluid velocity is faster because the actual computer operations are done on the C level and not on the Tcl level.\ 
\end{description}

\section{Current computational capabilities}
\label{sec:capabilities}
For simulations of blood flow using the OIF framework, we need to make a compromise between the accuracy of blood cell model and the computational capabilities. It is not reasonable to have too detailed model of the red blood cell with thousands of boundary points. From our experience it is sufficient to use several hundreds of IB points to get the proper elastic behavior. 

Our model is composed of two components, the fluid and the immersed objects. For both components a specific computational module is involved: For the fluid it is the lattice-Boltzman method with a fixed number of grid points, and for the immersed objects, it is a Newton equations integrator, with each immersed object discretized. In the LB method the simulation box is discretized by a regular lattice with $n_x, n_y, n_z$ being the size of the LB lattice in particular direction. That means that the number of unknowns in the LB method is proportional to $n_x\times n_y\times n_z$. The IB method is discretized by the triangular mesh so that the size of the problem depends on the size of the mesh $n_{ib}$ and on the number of immersed objects $n_{obj}$. In total, the number of unknowns in the IB method is proportional to $n_{obj} \times n_{ib}$.

The overall computational complexity thus scales with 
$$n_x \times n_y \times n_z +  n_{obj} \times n_{ib}.$$ 

From the previous expression one can make a clear picture of the interplay between the complexity levels on both components of the model. For our preliminary simulations we use a triangular mesh with 393 IB points.  For illustration, on a standard notebook with a single core one can simulate the movement 5-10 blood cells in a channel of dimension 200x30x30 and the results are available in minutes. The length of the simulation is about 100ms.

On a stronger machine with 16 cores, 1.6GHz each, and a GPU, one is capable of simulating 100 blood cells during 100ms time interval in a channel of dimension 300x200x50 and the results are available in hours. These figures are only for illustration.

Further improvements of the capabilities are readily possible by using stronger computational resources. The supercomputers containing several thousands of cores enable computation of hundreds or thousands of red blood cells in larger domains. 

\section{Conclusion}
Some publications already have appeared that use the OIF framework for simulations \cite{gusenbauer2013,cimrak11a,Cimrak2013}. Currently, the OIF framework undergoes further developments. We are working on implementation of affinity mechanisms, so that in the future it will be possible to model the adhesion of cells to surfaces. We also plan to make the visualization of OIF data more convenient by implementing automatic generation of VTK files that can be easily viewed, e.g. in ParaView.

There have been collaborations started with biologists on simulating biological processes. Our aim is to make the OIF framework easy to use and well documented so that a scientist with basic programming skills will be able to master the Tcl code necessary to set up a simulation.

\end{document}